\documentclass[twocolumn]{aastex701}
\usepackage{graphicx}  
\usepackage{amsmath}   
\usepackage{booktabs}  
\usepackage{multirow}  

\begin{document}

\title{Cool Embeddings: Predicting Galaxy Cluster Cooling Times in IllustrisTNG and TNG-Cluster with AstroCLIP}

\author{R.~Karam}
\email{}
\affiliation{Département de physique, Université de Montréal, C.P. 6128 Succ. Centre-ville, Montréal H3C 3J7, Canada}

\author{C.~L.~Rhea}
\email{carterrhea93@gmail.com}
\affiliation{Département de physique, Université de Montréal, C.P. 6128 Succ. Centre-ville, Montréal H3C 3J7, Canada}
\affiliation{Centre de recherche en astrophysique du Québec (CRAQ)}
\affiliation{Dragonfly Focused Research Organization, 150 Washington Avenue, Santa Fe, NM 87501, USA}

\author{J.~Hlavacek-Larrondo}
\email{}
\affiliation{Département de physique, Université de Montréal, C.P. 6128 Succ. Centre-ville, Montréal H3C 3J7, Canada}
\affiliation{Centre de recherche en astrophysique du Québec (CRAQ)}

\author{L.~Perreault-Levasseur}
\email{}
\affiliation{Département de physique, Université de Montréal, C.P. 6128 Succ. Centre-ville, Montréal H3C 3J7, Canada}
\affiliation{Centre de recherche en astrophysique du Québec (CRAQ)}

\author{M.~Prunier}
\email{}
\affiliation{Département de physique, Université de Montréal, C.P. 6128 Succ. Centre-ville, Montréal H3C 3J7, Canada}
\affiliation{Centre de recherche en astrophysique du Québec (CRAQ)}
\affiliation{Max-Planck-Institut für Astronomie, Königstuhl 17, D-69117 Heidelberg, Germany}

\author{A.~Alexandre}
\email{}
\affiliation{Département de physique, Université de Montréal, C.P. 6128 Succ. Centre-ville, Montréal H3C 3J7, Canada}

\author{H.~Choi}
\email{}
\affiliation{Département de physique, Université de Montréal, C.P. 6128 Succ. Centre-ville, Montréal H3C 3J7, Canada}
\affiliation{Department of Astronomy, University of Michigan, 1085 S. University Ave., Ann Arbor, MI 48109, USA}


\begin{abstract}
Foundation models trained on optical astronomical data have not previously been applied to X-ray cluster thermodynamics. We show that embeddings from \texttt{AstroCLIP}, a contrastive vision--language model trained on optical galaxy images and spectra, retain physically meaningful structure when applied to mock \textit{Chandra} X-ray images of simulated galaxy clusters --- a significant shift in both observational modality and physical scale. More specifically, we target the central cooling time ($t_{\mathrm{cool},0}$) of the intracluster medium, a key diagnostic of the thermodynamic state of galaxy clusters. However, measuring $t_{\mathrm{cool},0}$ has traditionally required X-ray spectroscopy with facilities such as \textit{Chandra} or \textit{XMM-Newton}, limiting its scalability. We present a machine learning framework that predicts $t_{\mathrm{cool},0}$ directly from X-ray images, applied to 711 projections from TNG300 and 993 from TNG-Cluster, both suites of the \textit{IllustrisTNG} simulations. We show that AstroCLIP embeddings form a structured latent space in which clusters with similar thermodynamic properties are naturally grouped, with no equivalent separation by halo mass. Building on this, we introduce a dot-product $k$-nearest neighbour attention model that retrieves semantically similar clusters from the training set and adaptively reweights their cooling times. This model achieves root-mean-square errors of 0.24 (TNG300) and 0.23 (TNG-Cluster) in $\log(t_{\mathrm{cool},0})$, and demonstrates strong out-of-distribution generalisation, outperforming a fine-tuned convolutional neural network trained on raw images (RMSE 0.50 versus 0.73). These results demonstrate that pretrained foundation model embeddings transfer across astrophysical domains and encode thermodynamic information sufficient for spectroscopic inference of cluster cooling properties --- offering a scalable path toward cool-core characterisation in all-sky surveys such as \textit{eROSITA}.
\end{abstract}




\section{Introduction}
Galaxy clusters are the most massive gravitationally bound systems in the Universe \citep[e.g.][]{sarazin_x-ray_1986}. These clusters emit strongly in X-ray 
with bolometric X-ray luminosities ranging from $10^{43}$ to $10^{46}$ erg s$^{-1}$ \citep{sarazin_x-ray_1986}. Typical clusters have radii of a few megaparsecs (Mpc) and total masses between $10^{14}$ and $10^{15} M_{\odot}$ \citep{sparke_galaxies_2007}. Their mass is primarily composed of dark matter (84\%), with the intracluster medium (ICM) accounting for 13\%, and galaxies making up the remaining 3\% \citep[e.g.,][]{mohr_properties_1999, vikhlinin_chandra_2006, umetsu_subaru_2009}.

The ICM, consisting of hot, ionized gas—primarily hydrogen and helium—represents the bulk of the baryonic mass in galaxy clusters. With typical temperatures of $\sim 10^7$ K and electron densities around $10^{-1}$-$10^{-3}$ cm$^{-3}$, the ICM emits predominantly via thermal bremsstrahlung radiation \citep{sarazin_x-ray_1986, binney_galactic_2008}.

Galaxy clusters are commonly classified according to the thermodynamic state of their central ICM, with the central cooling time ($t_{\mathrm{cool},0}$; the 0 refers to $z=0$) serving as a key diagnostic. Systems with short $t_{\mathrm{cool},0}$ (e.g., $<1$ Gyr) are identified as strong cool-core (SCC) clusters, while those with long $t_{\mathrm{cool},0}$ (e.g., $\geq 7.7$ Gyr) and flat X-ray surface brightness profiles are considered non-cool-core (NCC) clusters. An intermediate population with $t_{\mathrm{cool},0}$ values between these thresholds is classified as weak cool-core (WCC) clusters \citep{cavagnolo_intracluster_2009, hudson_what_2010}. These categories reflect distinct thermal histories, merger histories, and varying levels of feedback regulation. A detailed description of the $t_{\mathrm{cool},0}$ definition and classification scheme is provided in Section~\ref{sec:coolingtime}.

The thermodynamic state of a galaxy cluster’s core—classified as SCC, WCC, or NCC—is shaped primarily by its dynamical history. In particular, mergers can destroy cool cores, producing WCC or NCC systems. On the other hand, AGN feedback regulates gas cooling and determines whether a cool core can reform \citep{rasia_cool_2013, barnes_role_2018, mcnamara_heating_2007, fabian_observational_2012}.  However, recent work suggests that AGN feedback can also, under certain conditions, contribute to the destruction of cool cores \citep{lehle_heart_2024}.

AGN feedback is generally divided into two modes based on the accretion rate onto the central supermassive black hole (SMBH). At high accretion rates ($L_{\text{AGN}} > 0.01L_{\text{Edd}}$), the radiative or quasar mode dominates, injecting energy through winds and radiation \citep{wandel_accretion_1991, gallagher_active_2007, churazov_evolution_2005, hopkins_quasars_2010, hamann_quasar_2013}. At lower accretion rates, where the AGN luminosity is well below the Eddington limit (i.e., $L_{\text{AGN}} < 0.01L_{\text{Edd}}$), feedback proceeds via the kinetic mode. The majority of galaxy clusters are in this kinetic regime, where relativistic jets inflate X-ray cavities and drive shocks into the surrounding ICM \citep[e.g.,][]{churazov_supermassive_2005, croton_many_2006, fabian_observational_2012, heckman_coevolution_2014}.

To study these complex feedback processes, cosmological magnetohydrodynamical simulations have become indispensable. Large-scale projects such as IllustrisTNG simulate galaxy formation and evolution from $z = 127$ to the present day by solving coupled equations for gravity, magnetohydrodynamics (MHD), and with subgrid presciptions to model star formation, supermassive black hole seeding, feedback and feeding, and black hole feedback \citep{springel_first_2018, pakmor_improving_2016}. The TNG50, TNG100, and TNG300 simulations span increasing comoving volumes, offering a trade-off between spatial resolution and statistical sample size \citep[e.g.,][]{pillepich_first_2018, nelson_first_2019, nelson_illustris_2015, nelson_first_2018, nelson_first_2019, weinberger_supermassive_2017}. 
The typical mass of galaxy clusters M500c = 1e14-1e15 Msun $M_{500,c}=1\times10^{14} - 1\times 10^{15} M_{\odot}$. While these initial cosmological boxes contain a limited number of galaxy clusters specifically at very massive ones, TNG-Cluster, a spin-off project of IllustrisTNG consist of 352 cosmological zoom-ins of galaxy clusters. 
In this work, we utilize both the TNG300 and TNG-Cluster simulations \citep[see also][]{engelmann_tng-cluster_2023}.

On the observational side, the \textit{Chandra X-ray Observatory} \citep{graessle_iridium_2000}, with its sub-arcsecond angular resolution, has played a central role in characterizing the ICM, AGN feedback, and cool-core phenomena. More recently, the \textit{eROSITA} mission aboard the Spektr-RG observatory has begun delivering an unprecedented all-sky X-ray survey, dramatically expanding the available sample of galaxy clusters and providing new opportunities to study cluster thermodynamics and cool-core classification at scale \citep{predehl_erosita_2021, merloni_erosita_2012, bulbul_erosita_2021, truong_erosita_2021, ghirardini_characterizing_2022, klein_erosita_2021}. These rapidly growing datasets call for automated and scalable analysis methods, such as the machine learning (ML) framework we explore in this work.

ML provides scalable and automated tools for extracting physical insights from high-dimensional datasets \citep{lecun_deep_2015, alcorn_machine_2021, baron_machine_2019, ball_data_2010, fluke_surveying_2020, dobricic_machine_2023}. In the context of galaxy clusters, ML has been applied to tasks such as mass estimation and dynamical state classification using multiwavelength data \citep{ntampaka_deep_2017, ntampaka_machine_2019, ho_constraining_2019, green_classifying_2019, de_andra_automated_2021, chiu_machine_2022, armitage_machine_2019}. While observational studies like \citet{mcdonald_galaxy_2019} have established the importance of cool-core structure, ML-driven approaches for directly probing ICM cooling properties remain relatively underexplored. Recently, \citet{sadikov_galaxy_2025} applied unsupervised clustering, deep learning regression, and simulation-based inference to mock \textit{Chandra} X-ray images at $z=0$ snapshots in IllustrisTNG simulations, successfully predicting multiple cool-core metrics—including central $t_{\mathrm{cool},0}$, entropy excess, and concentration parameter—with high accuracy, thereby demonstrating the potential of ML for large upcoming X-ray surveys such as \textit{eROSITA}.

Contrastive learning (CL) is a powerful self-supervised ML technique that learns representations by comparing similar and dissimilar pairs of data \citep{hadsell_dimensionality_2006}. Without requiring labels, CL encourages the model to draw augmented views of the same input (positive pairs) closer in embedding space, while pushing apart representations of unrelated samples (negative pairs).
Modern implementations, such as the \emph{Simple Framework for Contrastive Learning of Visual Representations} (SimCLR; \citet{chen_simple_2020}), \emph{Momentum Contrast} (MoCo; \cite{he_momentum_2020}), and \emph{Bootstrap Your Own Latent} (BYOL; \cite{grill_bootstrap_2020}), have demonstrated that CL can produce representations rivaling or surpassing those learned through supervised training. By leveraging unlabeled data, CL enables robust and generalizable feature extraction, making it well-suited for downstream tasks such as classification and regression. In astronomy, CL has been successfully applied to galaxy morphology classification \citep{huertas-company_brief_2023} and self-supervised representation learning for astronomical images \citep{hayat_self-supervised_2021}. Recently, \citep{chadayammuri_ergo-ml_2026} explored the use of contrastive learning algorithms to infer galaxy cluster properties.

In this work, we apply AstroCLIP, the CL framework proposed by \citet{parker_astroclip_2023}, which is designed to learn physically meaningful representations from multiwavelength astronomical data without requiring labeled training samples. AstroCLIP extends the CLIP paradigm \citep{radford_learning_2021} to the astrophysical domain by jointly training image and spectroscopic encoders, enabling the learned embedding space to capture both morphological and spectral features of astronomical sources. By leveraging large, heterogeneous datasets, AstroCLIP produces general-purpose representations that can be fine-tuned for specific downstream tasks. Here, we adapt AstroCLIP to mock \textit{Chandra} X-ray observations of simulated galaxy clusters to train a regression model to estimate central $t_{\mathrm{cool},0}$. 
 
The remainder of this paper is organized as follows. Section~\ref{sec:Data} describes the simulation datasets and mock X-ray observations. Section~\ref{sec:methods} presents the AstroCLIP framework, regression models, and experimental setup. Section~\ref{sec:results} presents the results, followed by a discussion in Section~\ref{sec:discussion}. Finally, Section~\ref{sec:conclusion} summarizes our conclusions and outlines future directions.

\section{Data}\label{sec:Data}

\subsection{Simulated Data: IllustrisTNG and TNG-Cluster}
Historically, galaxy clusters have often been categorized into discrete classes: SCC, WCC, and NCC based on thresholds in their central $t_{\mathrm{cool},0}$ values \citep[e.g.,][]{predehl_erosita_2021, merloni_erosita_2012, bulbul_erosita_2021}. While this classification scheme has proven useful, it compresses the inherently continuous nature of $t_{\mathrm{cool},0}$ into broad bins, causing clusters with significantly different physical states to be grouped together. In this work, we treat the problem as a regression task, predicting $t_{\mathrm{cool},0}$ directly from cluster X-ray images. This approach preserves the continuous nature of the parameter, enabling both the recovery of the traditional categories and the extraction of more precise physical information. The ultimate goal of this study is to develop a method capable of estimating $t_{\mathrm{cool},0}$ from imaging data alone, with sufficient accuracy to provide reliable values on the first attempt, thereby bypassing the need for time-intensive spectral analysis for every cluster.

To develop and evaluate models for galaxy cluster characterization, we require large, homogeneous samples of clusters spanning a wide range of masses, redshifts, and physical conditions. However, current observational surveys lack the volume, resolution, and consistency to provide such datasets. To address this, we turn to cosmological MHD simulations.

The IllustrisTNG suite of cosmological simulations investigates the formation and evolution of galaxies and large-scale structure within a Lambda Cold Dark Matter (\(\Lambda\)CDM) framework; the standard cosmological model in which structure forms hierarchically under the influence of cold dark matter and accelerated expansion driven by a cosmological constant \(\Lambda\) \citep{planck_collaboration_planck_2016, eisenstein_detection_2005}. These simulations span a range of volumes and resolutions, with the flagship runs TNG50, TNG100, and TNG300, encompassing comoving box sizes of respectively 50, 100, and 300 Mpc \citep{pakmor_improving_2016, springel_first_2018}. Each simulation evolves a representative volume of the universe from redshift \( z = 127 \) to \( z = 0 \), tracking the dynamics of dark matter, gas, stars, and SMBHs by solving the coupled equations of gravity and magnetohydrodynamics (MHD) using the \textsc{AREPO} code \citep{springel_e_2010}. The simulations adopt the same cosmology as the rest of the TNG suite, based on \citet{planck_collaboration_planck_2016}. The physical model includes radiative gas cooling, star formation, metal enrichment, stellar and AGN feedback, and SMBH dynamics, identical to the model described in prior TNG publications \citep[e.g.,][]{pillepich_first_2018, weinberger_supermassive_2017, vogelsberger_model_2013}.

A key feature of the TNG feedback model is the dual-mode AGN prescription: SMBHs inject energy into their surroundings either thermally (during high-accretion, or ``quasar'' mode) or kinetically (during low-accretion, or ``wind'' mode), with the transition governed by the Eddington ratio \citep[e.g.,][]{weinberger_supermassive_2017}. This framework allows the simulation to self-consistently regulate cluster-scale gas content and entropy profiles.The largest volumes TNG300 and TNG-Cluster achieve a baryonic mass resolution near \( 10^{7} \, M_\odot \), allowing for detailed studies of the ICM and feedback physics in a cosmological context.

TNG300 is one of the largest-volume simulations in the IllustrisTNG suite, covering a comoving volume of (205 Mpc $h^{-1}$)$^{3}$, approximately 300 comoving Mpc on a side. Its large spatial coverage makes it particularly well-suited for statistical studies of galaxy clusters and their role in the cosmic web, while still maintaining sufficient resolution to capture key physical processes such as star formation, AGN feedback, and ICM evolution \citep{pillepich_first_2018, nelson_first_2018, weinberger_supermassive_2017}. 
The moving-mesh MHD code \textsc{AREPO}, which adaptively refines the mesh to maintain high resolution in regions of high density \citep{pakmor_improving_2016}. This enables to accurately model the cluster cores thermodynamic and kineamtic evolution in a full cosmological context. . In this study, we utilize a subset of the 280 most massive halos at \( z = 0 \), selected based on their virial masses, corresponding to a threshold of approximately \( M_{\mathrm{vir}} \gtrsim 3.2 \times 10^{14}\, M_{\odot} \) \citep{springel_first_2018}.

TNG-Cluster \citep{nelson_tng-cluster_2024} is an extension of the IllustrisTNG project, comprising 352 high-resolution zoom-in simulations of massive galaxy clusters with \( M_{\mathrm{500c}} \gtrsim 10^{14} \, M_{\odot} \). Validation studies have shown that global cluster properties in TNG-Cluster are consistent with observational benchmarks, broadly reproducing the observed fraction of cool-core and non–cool-core systems, radial thermodynamic profiles, and key scaling relations \citep{nelson_tng-cluster_2024, ayromlou_tng-cluster_2024, lee_tng-cluster_2024, truong_tng-cluster_2024, lehle_heart_2024, rohr_tng-cluster_2024}. The clusters also reproduce realistic small scale X-ray structures such as AGN-driven cavities and shocks  \citep{prunier_x-ray_2025, prunier_x-ray_2025-1, prunier_x-ray_2025-2}
). As such, TNG-Cluster serves as a robust testbed for ML applications aimed at predicting cluster properties from mock observations.

For TNG300 and TNG-Cluster, we restrict our analysis to redshift \( z = 0 \), where the relevant cluster-scale quantities—such as virial mass, radius, and temperature—are well-defined and directly comparable to low-redshift observational benchmarks.

\subsubsection{$t_{\mathrm{cool},0}$ and CC Classification}
\label{sec:coolingtime}

For this study, we use pre-computed $t_{\mathrm{cool},0}$ values and classification flags provided in the IllustrisTNG database, following the methodology of \citet{lehle_heart_2024}. We use the following quantities: the central 
$t_{\mathrm{cool},0}$ values, measured within a 3D radius of \(0.012 \, R_{500c}\) by \citet{lehle_heart_2024} and consistent with prior observational and simulation-based studies \citep{mcdonald_galaxy_2019}; and the associated classification flags, which label each cluster as a SCC, WCC or NCC system. 

Although we use the tabulated values directly, we define here $t_{\mathrm{cool}}(r)$ as the ratio of the gas's specific thermal energy to its radiative energy loss rate:
\begin{equation}
t_{\mathrm{cool}}(r) \approx
\frac{3}{2}\,
\frac{(n_e+n_i)\, k_B T(r)}
{n_e n_i \Lambda(T,Z)} .
\end{equation}
where \( n_e \) is the electron number density, \( T \) is the temperature, \( k_B \) is the Boltzmann constant, and \( \Lambda(T, Z) \) is the cooling function, which depends on both temperature and metallicity \citep[e.g.,][]{sarazin_x-ray_1986}.

Following the classification scheme of \citet{lehle_heart_2024}, a cluster is labeled as SCC if its central $t_{\mathrm{cool},0}$ is less than 1 Gyr, as WCC if it lies between 1 and 7.7 Gyr, and as NCC if \( t_{\mathrm{cool},0} \geq 7.7 \) Gyr. To ensure consistent matching between halo properties and cooling classifications, we retrieve the Halo IDs, $t_{\mathrm{cool}}s$, and classification flags at this snapshot. In TNG-Cluster and TNG300, additional halo-level properties (e.g., total mass, $R_{500c}$, temperature profiles) are readily available for further validation and downstream analysis. While we do not use these class labels during training, they provide a useful reference for interpreting the distribution of clusters in embedding space and comparing our results with previous studies.

\subsection{Finding the Cluster Center}
\label{sec:center}
Defining a robust cluster center is a necessary preprocessing step to ensure that the mock X-ray images are consistently aligned; without it, variations in centering could introduce spurious differences across clusters and degrade the model’s ability to learn physically meaningful correlations.


Given our focus on X-ray morphology, which helps distinguish CC and NCC clusters, we adopt the method of \citet{sadikov_galaxy_2025}, which consists of an iterative, multi-resolution density mapping technique to identify the peak of the gas density distribution. This method provides a physically motivated estimate of the cluster core and ensures robust identification even in disturbed systems.

We bin the gas particle positions into a three-dimensional (3D) histogram, where each bin is weighted by the gas particle density; we note that we exclude cold gas ($T_{gas}<1\times 10^6$K. In addition to the weighted histogram, a simple count histogram is generated to compute a mean density in each bin if desired. To suppress small-scale noise and enhance the most prominent density peak, the 3D histogram is convolved with a Gaussian filter using the \texttt{gaussian\_filter} function from \texttt{scipy.ndimage}. The smoothing scale ($\sigma$) is defined in kpc and adjusted according to the histogram bin size. We apply smoothing scales of 250, 50, and 5\,kpc in successive iterations to progressively capture finer-scale structure. Large smoothing scales suppress small-scale substructure, enabling the identification of the global density maximum, while smaller scales reveal finer features for a more precise localization of the center. 

The process is carried out iteratively at three resolution levels, with bin widths of 10, 5, and 1\,kpc used to compute particle densities at each corresponding smoothing step. After each iteration, we restrict the dataset to focus on the most relevant region: we retain only particles within $\pm 3\sigma$ of the identified center along each coordinate, where $\sigma$ matches the current Gaussian smoothing scale. This filtering ensures that subsequent iterations focus on the densest core while mitigating contamination from substructures.

Although the algorithm autonomously converges on a center, we emphasize the importance of manual verification. At each iteration, we project the filtered particle distribution onto the \textit{X-Y} plane and generate a 2D histogram of the density distribution, clearly marking the estimated center. These diagnostic plots help track whether the densest region is being consistently identified throughout the iterative process and reveal potential anomalies such as significant substructures or poor convergence due to insufficient smoothing.


\subsection{Mock Chandra X-ray Images}
To connect our method to real observations, we generate mock \textit{Chandra} X-ray images of the simulated clusters. 
Simulations provide direct access to 3-D thermodynamic quantities such as density, temperature, and $t_{\mathrm{cool}}$s, for our study we will not directly use these simulated quantites but rather we want to mimic an observational study who are inferring tcool from X-ray images with all the systematics involved'. To that purpose, we will generate 2D mock Chandra images to train our model; in practice, $t_\mathrm{cool}$ is inferred from high-resolution X-ray spectroscopy. 
Our goal, however, is to develop a scalable model for upcoming large-area surveys, which requires realistic 2D images that reflect the limitations of real instruments.

For each target cluster ID, we extract the gas particle data—including positions and densities—from snapshot 99 of the IllustrisTNG simulation using the \texttt{illustris\_python} module \citep{nelson_illustris_2015}. Since the simulation enforces periodic boundary conditions, we apply a wrapping procedure via \texttt{handle\_box\_edges} to correctly remap particles near the edges of the simulation box. This function shifts particle coordinates so that the cluster is fully contained within a contiguous subvolume, ensuring spatial continuity and preventing artificial discontinuities in the gas density field.

Once we identify the 3D cluster center (Section~\ref{sec:center}), we generate mock ACIS-I \textit{Chandra} X-ray images using the \texttt{pyXSIM} \citep{zuhone_pyxsim_2016} and \texttt{SOXS} \citep{zuhone_soxs_2023} software packages. 
Photons in the 0.5–10.0 keV energy band are generated for each cluster within a cubic region of $\pm 1 \times R_{500}$ centered on this position. 
We include all gas cells within this volume, not only those gravitationally bound to the central galaxy or identified by a friends-of-friends algorithm. 
The only excluded component is star-forming gas, defined as gas cells with net cooling rates greater than zero (i.e., experiencing net heating).
This gas is excluded because it belongs to the multiphase star-formation model implemented in the simulation and is not expected to contribute significantly to the diffuse X-ray emission of the intracluster medium. Including it could therefore bias estimates of the gas cooling properties relevant to this work.

For each gas cell, we generate a mock X-ray spectrum based on its density, temperature, and metallicity. We assume a single-temperature APEC model \citep{smith_collisional_2001} using the simulation's abundance ratios, and we apply Galactic absorption with a hydrogen column density of $4 \times 10^{20} \, \mathrm{cm}^{-2}$. The spectra from all gas cells within the cubic core region are sampled to yield a large sample of photons per cluster and line of sight.

These initial photon samples are then passed to \texttt{SOXS}, which simulates observed Chandra events by projecting the photons onto the detector plane and convolving them with an instrument model for the ACIS-I detector. We use Cycle 19 instrumental response files and the on-axis point-spread function (PSF) for ACIS. To match the ACIS broad energy band, we restrict the final energy range to 0.5–7.0 keV.

To ensure realistic observational conditions, we include X-ray emission from satellite galaxies within the same halo, as well as instrumental and cosmic X-ray backgrounds and the Milky Way foreground. Each cluster is simulated along three orthogonal projections (aligned with the simulation's X, Y, and Z axes), and we adopt an exposure time of 200 ks. 

After photon simulation, we perform background subtraction and remove point sources using the \texttt{CIAO} wavelet detection algorithm (\texttt{wavdetect}) with scales set to 2.0 and 4.0. This step is crucial, as a bright point source near the cluster core could dominate the emission, leading to incorrect identification of the X-ray center. Moreover, removing point sources ensures that downstream analyses and machine learning algorithms can derive meaningful information from the cluster emission without contamination. Detected point sources are masked using elliptical regions and replaced with interpolated values using the \texttt{dmfilth} tool to preserve the surrounding diffuse emission. The resulting images are then rebinned to a resolution of $252 \times 252$ pixels (as AstroCLIP requires dimensions that are multiples of 12) and min–max normalized to prepare them for input into our ML pipeline.

In total, we began with 237 clusters from TNG300 and 331 clusters from TNG-Cluster. 
For each cluster, we identified the center and generated three mock \textit{Chandra} images by projecting along the $x$, $y$, and $z$ axes. 
This procedure yields 711 images for TNG300 and 993 images for TNG-Cluster, which we use as inputs to our embedding and regression experiments.
\begin{figure*}
    \centering
    \includegraphics[width=0.95\textwidth]{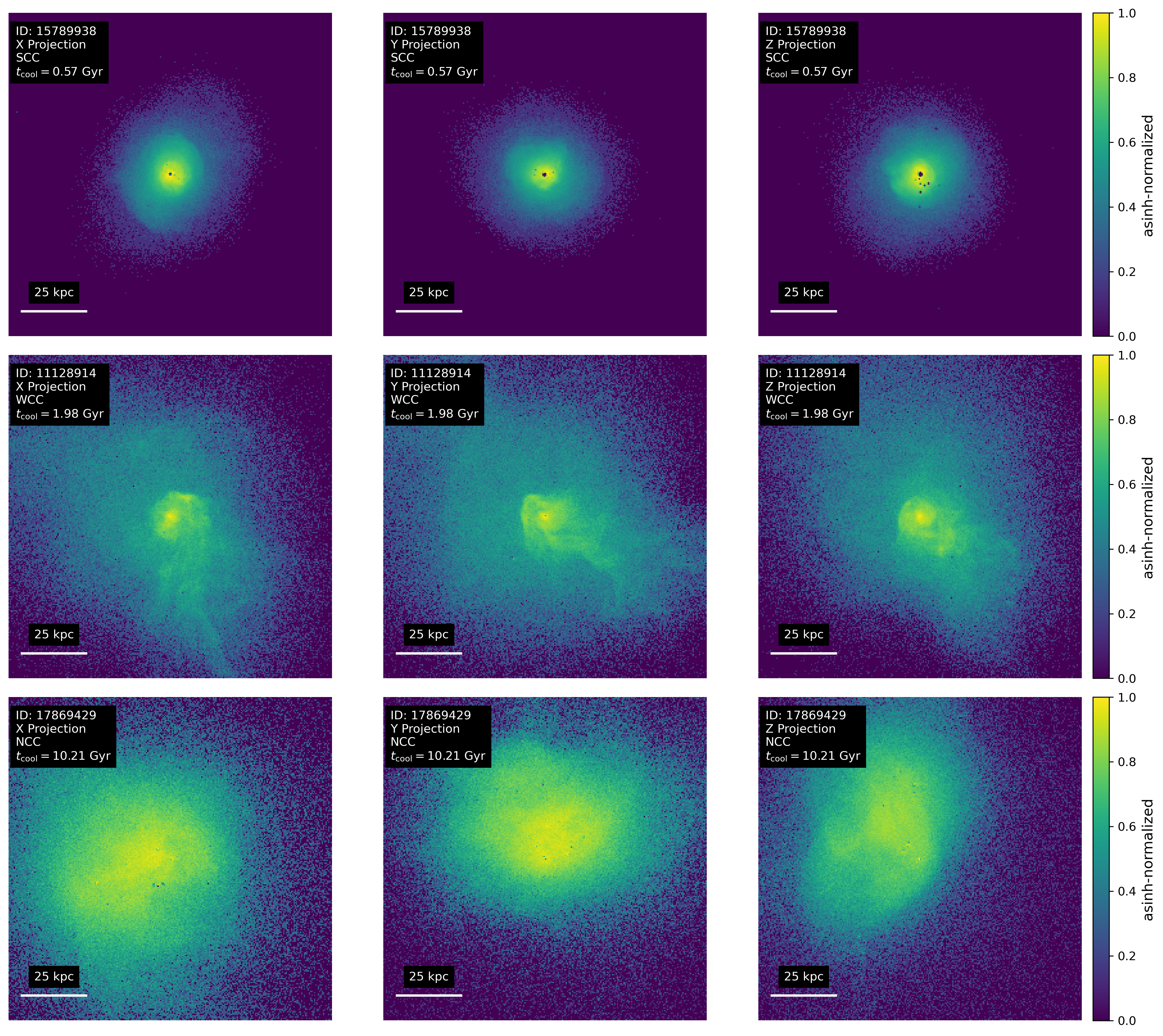}
    \caption{Example X, Y, and Z projections of three TNG-Cluster halos, each representing a different cool-core class (SCC, WCC, NCC). 
        Pixel values use an asinh stretch with shared colour normalization. Qualitatively the X-ray morphologies of the cores are distinct; centrally peaked and concentrated for the SCC, more extended with a strong sloshing spiral pattern for the WCC and disturbed non-symetrical morphology for the NCC. }
    \label{fig:tngcluster_tcool}
\end{figure*}


\section{Methods} \label{sec:methods}

Our goal is to train an algorithm capable of directly estimating $t_\mathrm{cool}$, from galaxy cluster images without relying on spectroscopic information. 
This is a challenging task, since $t_\mathrm{cool}$ is a thermodynamic quantity traditionally derived from X-ray spectroscopy, whereas here we aim to infer it from imaging data alone. 
To this so, we leverage \texttt{AstroCLIP} to embed clusters into a latent space that preserves this physically meaningful structure.

\subsection{AstroCLIP Embeddings}

AstroCLIP is a contrastive vision-language foundation model introduced by \citet{parker_astroclip_2023} and developed by the Polymathic AI collaboration. It builds on the CLIP framework \citep{radford_learning_2021} and uses transformer-based encoders: a Vision Transformer (ViT) for galaxy images and a standard transformer for spectra. The model is trained to embed both modalities into a shared latent space.

ViTs operate by dividing images into fixed-size patches and modeling the relationships between these patches using self-attention mechanisms \citep{dosovitskiy_image_2021}, which compute each patch's representation as a weighted combination of all others. The weights are learned dynamically and reflect the relevance of each patch to the one being updated, enabling the model to capture both local and global contextual information.

The training is conducted in a self-supervised manner, meaning that training labels are constructed during training by the model from paired data rather than being explicitly provided prior to training. It does so using a contrastive loss applied to paired observations from the Legacy Imaging Survey \citep{dey_overview_2019} and the Dark Energy Spectroscopic Instrument (DESI) \citep{collaboration_desi_2016}. The objective of the contrastive loss allows for matched pairs—whether image–image, spectrum–spectrum, or image–spectrum—to lie close together in latent space, while pushing apart mismatched pairs.

To construct positive training pairs, AstroCLIP applies strong domain-specific augmentations to individual galaxy observations, such as random crops, rotations, and colour jittering for images, or spectral masking and jittering for spectra, ensuring that semantically similar inputs are mapped close together in the embedding space \citep{parker_astroclip_2023}.

AstroCLIP embeddings have been shown to support diverse astrophysical tasks, including photometric redshift estimation, galaxy morphology classification, and stellar mass prediction, with performance competitive with or exceeding that of fully supervised baselines \citep{parker_astroclip_2023}. Foundation models like AstroCLIP allow for generalization across instruments and tasks, making them especially effective in low-label or simulation-to-observation transfer scenarios. In this work, we use the 1024-dimensional embeddings generated by the ViT encoder as fixed representations for downstream regression tasks.

In our study, we test the domain generalization capabilities of AstroCLIP by applying its vision encoder to a new astrophysical setting: mock X-ray images of galaxy clusters. Although AstroCLIP was trained on optical galaxy-scale data, we evaluate whether its representations remain informative for a distinct physical regime and observational modality. We generate 1024-dimensional embeddings for $252 \times 252$ single-channel mock X-ray images, with each cluster projected along the X, Y, and Z axes. These grayscale X-ray images differ significantly from the RGB optical galaxy images used in pretraining, representing both a spectral and morphological domain shift. The resulting three images are embedded independently, without additional normalization or preprocessing. This procedure yields 711 image embeddings for TNG300 and 993 for TNG-Cluster. For experiments on the merged dataset, we concatenate the embeddings from both simulations.

\subsection{Regression Setup and Model Architectures}

We evaluate four supervised regression models on AstroCLIP embeddings, each trained to predict the $t_{\mathrm{cool}}$ in log space. The models differ in complexity, inductive bias, and how they incorporate contextual information from the embedding space.

All models are trained using 80\% of the available data, with 20\% held out for testing. The split is performed at the cluster level to prevent data leakage: all three projections of a given cluster are assigned to either the training or testing set, but never both.

We test the following four regression models: multi-layer perceptron, ResNet, Singlle layer attention, and dot KNN attention. For the MLP, ResNet, and Single-Layer Attention (Attn), we use the Adam optimizer together with the Huber loss; a piecewise function that behaves like L2 (squared error) for small residuals and like L1 (absolute error) for large residuals. A grid search is performed over learning rates ${10^{-3},\ 5\times10^{-4},\ 10^{-4}}$ and weight decay values ${10^{-4},\ 10^{-5},\ 10^{-6}}$, with early stopping (patience = 10 epochs) based on validation root mean square error (RMSE) . Each model is trained for up to 75 epochs with a batch size of 32.

For the fourth model, Dot-Product Attention over kNNs (kNN-Attn), we fix the learning rate to $10^{-3}$ and omit weight decay. Because the model contains relatively few trainable parameters and relies primarily on information retrieved from neighboring samples in embedding space, preliminary experiments showed little sensitivity to weight decay. A learning rate of $10^{-3}$ provided stable convergence across datasets and was therefore used throughout.

\subsubsection{Multi-Layer Perceptron: MLP}

As a baseline model, we use a standard fully connected MLP, a widely adopted architecture for regression tasks on fixed-length feature vectors \citep{goodfellow_deep_2016}. The model takes as input a 1024-dimensional AstroCLIP embedding derived from a single mock X-ray image projection. It consists of three hidden layers with 512, 256, and 128 units, each followed by a ReLU activation and dropout (rate = 0.1). A final linear layer outputs the scalar prediction of $\log(t_{\mathrm{cool}})$.

\subsubsection{Residual network: ResNet}

To enable deeper representation learning while mitigating vanishing gradients, we implement a custom residual network inspired by \citet{sadikov_galaxy_2025}. The input embedding is first projected to 512 dimensions via a linear layer, followed by batch normalization and a LeakyReLU activation. This is followed by three stacked residual blocks operating at 512 dimensions. The representation is then progressively reduced in dimensionality through linear layers interleaved with LeakyReLU activations and additional residual blocks at 256 and 128 dimensions. A dropout layer (rate = 0.4) is applied before the final linear output layer, which produces the predicted $\log(t_{\mathrm{cool}})$.

\subsubsection{Attn}

Motivated by the attention-based architecture of AstroCLIP, we employ a lightweight attention mechanism to perform feature selection over the 1024-dimensional embedding $\mathbf{x} \in \mathbb{R}^{1024}$. A learned weight vector $\mathbf{w} \in \mathbb{R}^{1024}$ computes a scalar gate:
\[
\alpha = \sigma(\mathbf{w}^\top \mathbf{x}),
\]
where $\sigma$ denotes the sigmoid activation. The embedding is then reweighted via $\alpha \cdot \mathbf{x}$ and passed through a two-layer multilayer perceptron (MLP) with hidden dimensions 256 and 128, LeakyReLU activations, and dropout.

\subsubsection{kNN-Attn}

This model combines nonparametric neighbor retrieval with parametric attention, by leveraging the expressiveness of attention mechanisms to integrate contextual information from nearby embeddings. For a given input embedding $\mathbf{x} \in \mathbb{R}^{1024}$, which serves as the \textit{query}, the model retrieves its $k$ nearest neighbors $\{\mathbf{x}_i\}_{i=1}^k$ from the training set based on Euclidean distance between Astroclip embedding vectors. These neighbors act as the \textit{keys}, and their associated $t_{\mathrm{cool}}$s $\{y_i\}_{i=1}^k$ serve as the \textit{values}. The architecture projects the input $\mathbf{x}$ (query), retrieved embeddings $\mathbf{x}_i$ (keys), and their associated $t_{\mathrm{cool}}$s $y_i$ (values) into a shared latent space of dimension $d=128$ via learned linear projections:
\[
\mathbf{q} = W_q \mathbf{x}, \quad 
\mathbf{K} = [W_k \mathbf{x}_1, \dots, W_k \mathbf{x}_k], \quad 
\mathbf{V} = [W_v y_1, \dots, W_v y_k],
\]
where $W_q, W_k \in \mathbb{R}^{d \times 1024}$ and $W_v \in \mathbb{R}^{d \times 1}$.

Attention weights are computed using scaled dot-product attention:
\[
\mathbf{a} = \mathrm{softmax}\left( \frac{\mathbf{q} \mathbf{K}^\top}{\sqrt{d}} \right)
\]
The final attended prediction vector is obtained as:
\[
\hat{\mathbf{z}} = \mathbf{a} \cdot \mathbf{V},
\]
and the scalar output is:
\[
\hat{y} = W_o \hat{\mathbf{z}} + b_o,
\]
with $W_o \in \mathbb{R}^{1 \times d}$ and $b_o \in \mathbb{R}$.

\section{Results}\label{sec:results}
\begin{figure*}[t]
    \centering
    \includegraphics[width=\linewidth]{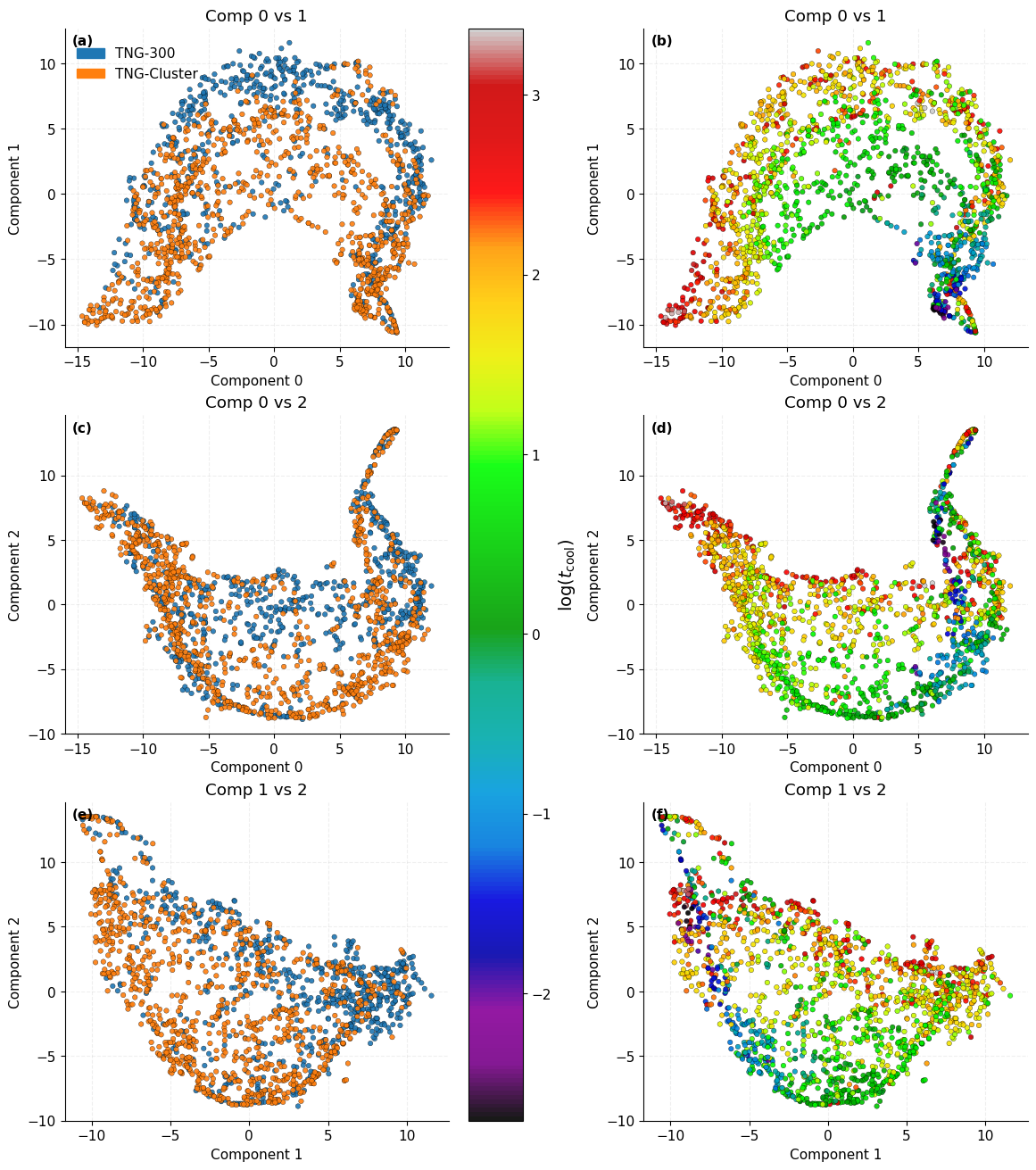}
    \caption[t-SNE projections of AstroCLIP embeddings]{
        t-SNE projections of AstroCLIP embeddings for galaxy clusters from TNG300 and TNG-Cluster. 
        Each row shows a different 2D projection of components: (0,1), (0,2), and (1,2). 
        In the left column, points are coloured by dataset (blue: TNG300, orange: TNG-Cluster), revealing that clusters from the two simulations occupy overlapping yet distinct subregions of the embedding space. 
        In the right column, we plot the same t-SNE plots, but points are coloured by $\log(t_{\mathrm{cool}})$, illustrating that AstroCLIP captures physically meaningful structure correlated with thermodynamic properties of the ICM.
    }
    \label{fig:tsne}
\end{figure*}
\begin{figure*}[t]
    \centering
    \includegraphics[width=\linewidth]{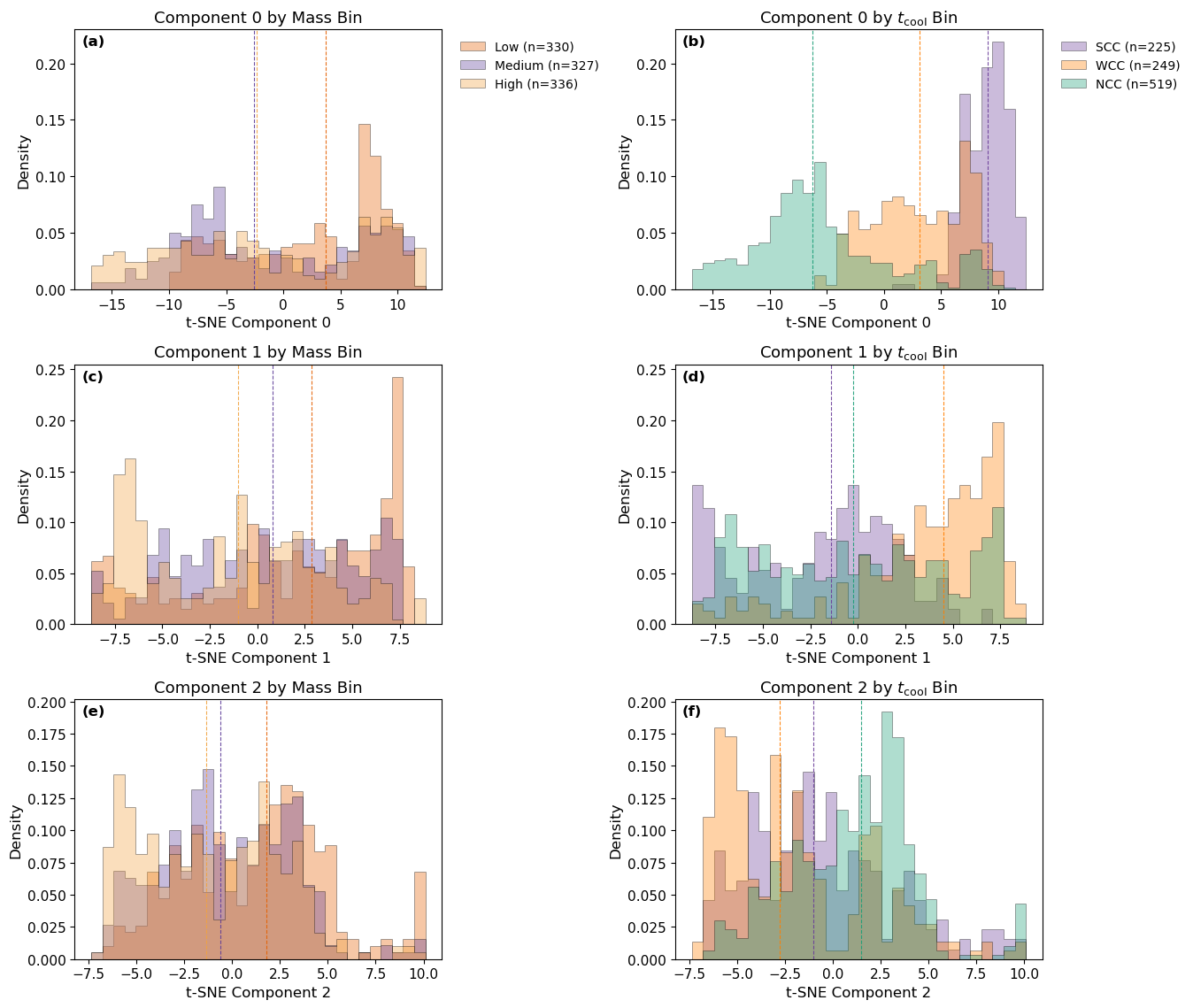}
    \caption[Mass and $t_{\mathrm{cool}}$ distributions in TNG-Cluster t-SNE space]{
        Distribution of t-SNE components (0, 1, and 2) of AstroCLIP embeddings for the TNG-Cluster sample, separated by mass bin (left panels) and $t_\mathrm{cool}$ category (right panels). 
        Mass bins are defined by the 33rd and 66th percentiles of $M_{500c}$, while $t_\mathrm{cool}$ categories are classified as SCC, WCC, or NCC following standard thresholds. 
        Mass bins show substantial overlap across all components, indicating no clear separation by mass in the embedding space. 
        In contrast, $t_\mathrm{cool}$ categories—particularly SCC and WCC—exhibit more localized distributions in components 0 and 1, whereas NCC clusters are more broadly dispersed. 
        Vertical dashed lines indicate the median of each distribution (by mass bin in the left panels, and by $t_{\mathrm{cool}}$ category in the right panels), where density corresponds to the number of clusters per bin.
    }
    \label{fig:tng_cluster_massdist}
\end{figure*}
\begin{figure*}[t]
    \centering
    \includegraphics[width=\linewidth]{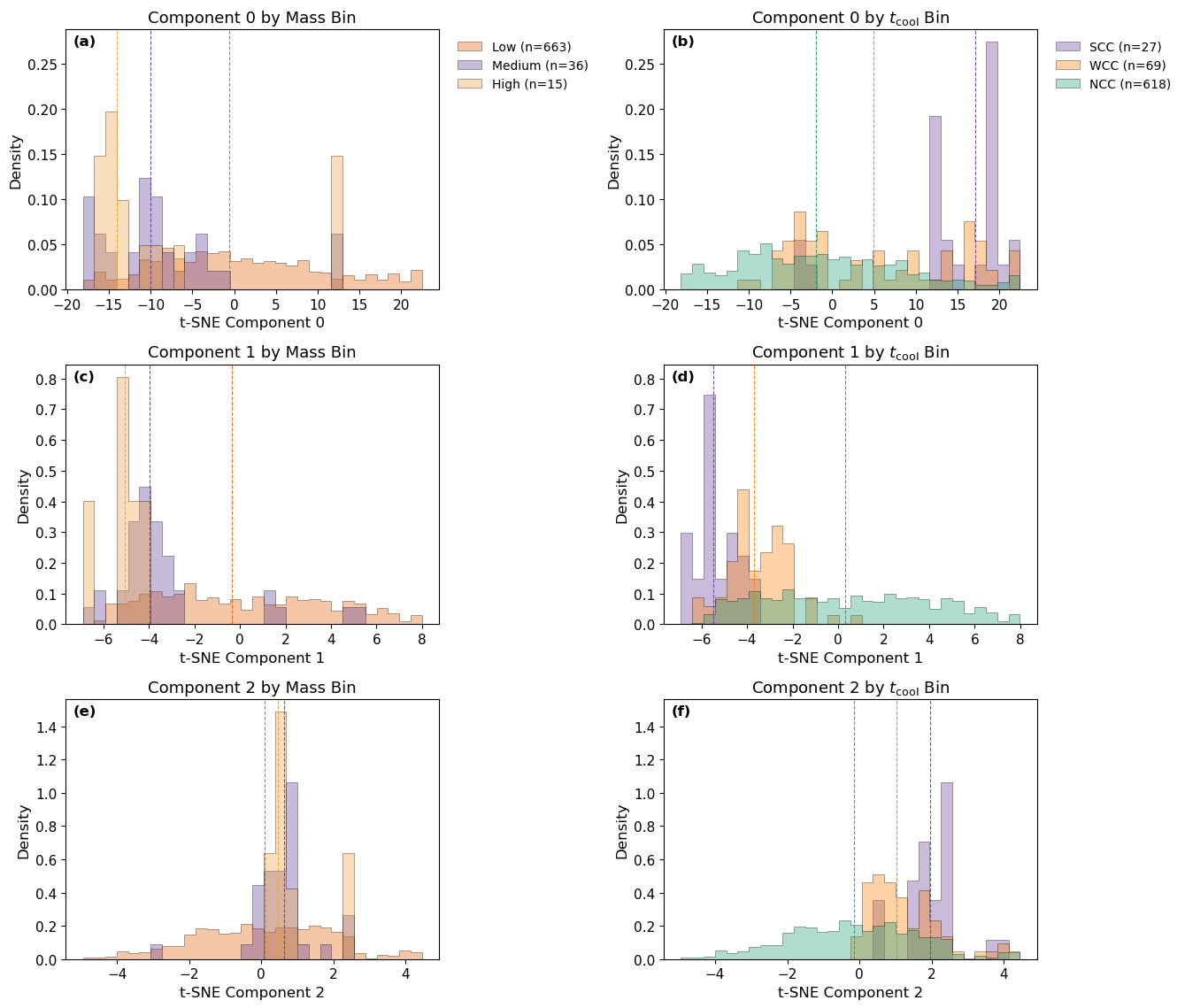}
    \caption[Mass and $t_{\mathrm{cool}}$ distributions in TNG300 t-SNE space]{
        Distributions of the three t-SNE components (0, 1, and 2) of AstroCLIP embeddings for the TNG300 sample, separated by cluster mass (left column) and $t_{\mathrm{cool}}$ category (right column). 
        Mass bins are defined as low (\(n=663\)), medium (\(n=36\)), and high (\(n=15\)), while $t_{\mathrm{cool}}$ categories are defined as SCC (\(n=27\)), WCC (\(n=69\)), and NCC (\(n=618\)). 
        Vertical dashed lines indicate the median value of each distribution (by mass bin in the left panels, and by $t_{\mathrm{cool}}$ category in the right panels).
    }
    \label{fig:tng_300_massdist}
\end{figure*}

The best-performing hyperparameter configurations and corresponding validation RMSE scores are summarized in Table~\ref{tab:best_hparams}.
\begin{table*}[t]
\centering
\caption{Best hyperparameters where applicable and validation RMSE for each model across datasets. Bold values indicate the best RMSE in each dataset. For kNN-Attn, we report the best number of neighbors $k$.}
\label{tab:best_hparams}
\begin{tabular}{lcccc}
\toprule
\textbf{Dataset} & \textbf{Model} & \textbf{Best LR / $k$} & \textbf{Best WD} & \textbf{Validation RMSE} \\
\midrule
\multirow{4}{*}{TNG300}
  & MLP        & $5\times10^{-4}$ & $10^{-4}$  & 0.43 \\
  & ResNet     & $5\times10^{-4}$ & $10^{-4}$  & 0.49 \\
  & Attn       & $5\times10^{-4}$ & $10^{-4}$  & 0.43 \\
  & kNN-Attn   & $k=10$           & ---        & \textbf{0.24} \\
\midrule
\multirow{4}{*}{TNG-Cluster}
  & MLP        & $10^{-4}$        & $10^{-4}$  & 0.40 \\
  & ResNet     & $10^{-4}$        & $10^{-5}$  & 0.45 \\
  & Attn       & $10^{-4}$        & $10^{-6}$  & 0.38 \\
  & kNN-Attn   & $k=15$           & ---        & \textbf{0.23} \\
\midrule
\multirow{4}{*}{Merged}
  & MLP        & $10^{-4}$        & $10^{-6}$  & 0.75 \\
  & ResNet     & $10^{-4}$        & $10^{-4}$  & 0.76 \\
  & Attn       & $10^{-4}$        & $10^{-4}$  & 0.74 \\
  & kNN-Attn   & $k=15$           & ---        & \textbf{0.59} \\
\bottomrule
\end{tabular}
\end{table*}

The three embedding-based regressorsachieve comparable performance across both simulations, with validation RMSEs ranging from 0.38 to 0.49. Increasing model complexity beyond a simple MLP yields only modest improvements, suggesting that the AstroCLIP embeddings are already highly informative and require relatively little additional feature learning. In contrast, incorporating local neighborhood information through the kNN-Attn model produces the largest performance gains, consistently achieving the lowest RMSE across all datasets. This indicates that exploiting the local structure of the embedding space is more beneficial than increasing the capacity of the regression model alone.

\begin{figure}[t]
    \centering
    \includegraphics[width=\linewidth]{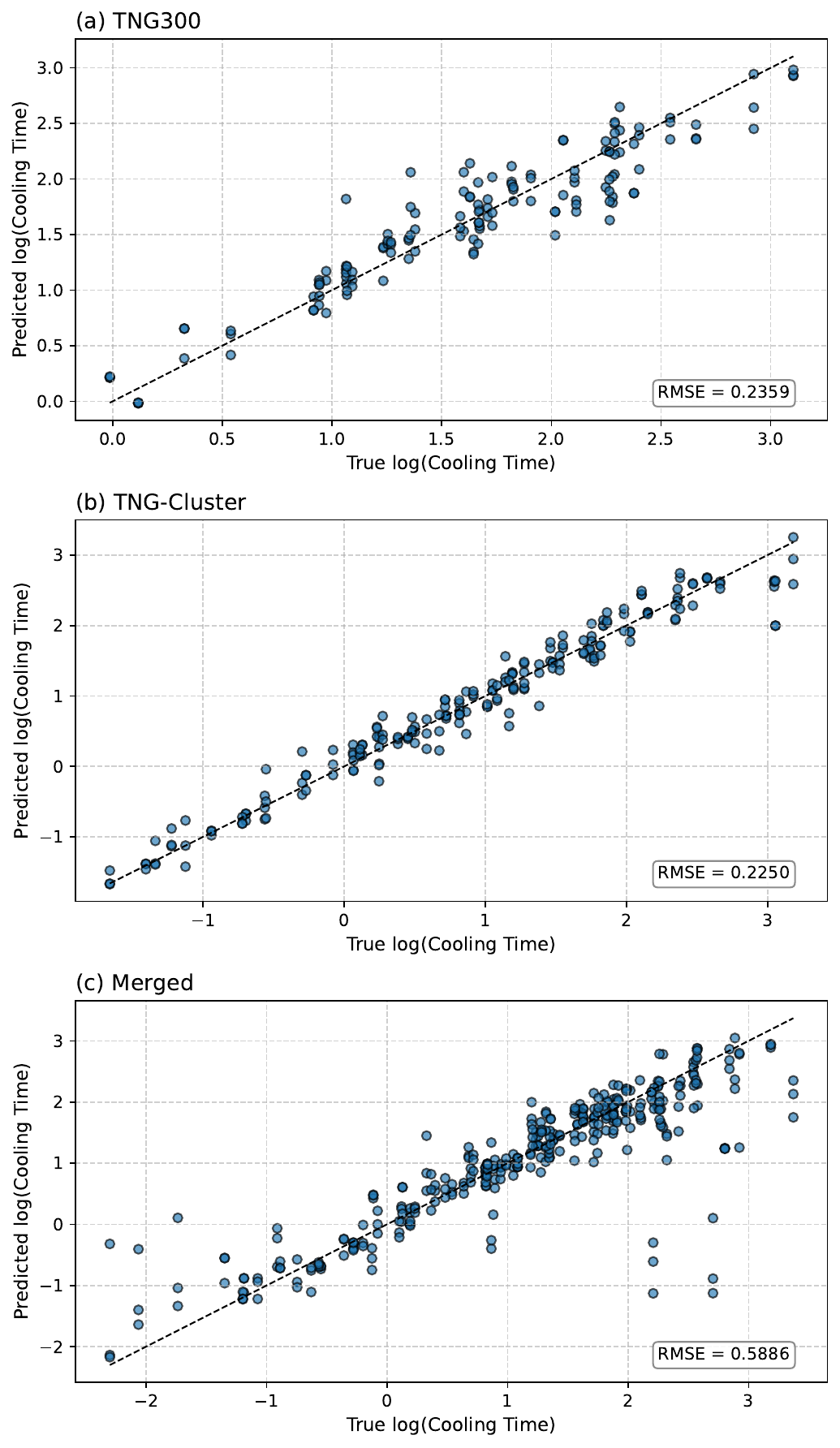}
    \caption[Predicted vs.\ true $\log(t_{\mathrm{cool}})$ for the kNN-Attn]{
        Predicted vs.\ true $\log(t_{\mathrm{cool}})$ for the kNN-Attn model on TNG300, TNG-Cluster, and the merged set. 
        The dashed line indicates the 1:1 reference.
    }
    \label{fig:dotattn}
\end{figure}

The kNN-Attn model achieves an RMSE of 0.503, compared to 0.729 for the CNN.
\begin{figure}[t]
    \centering
    \includegraphics[width=\linewidth]{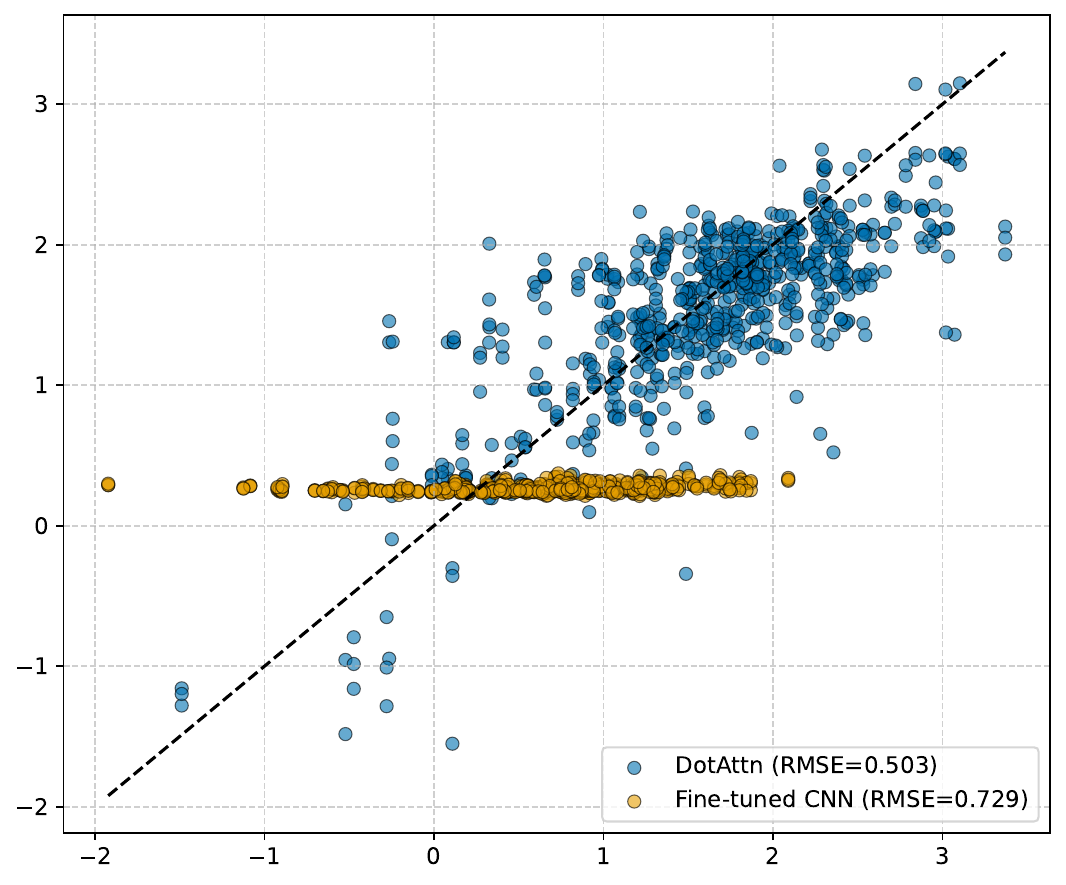}
    \caption{Predicted versus true $\log(t_{\mathrm{cool}})$ for the fine-tuned CNN and kNN-Attn models when evaluated on the TNG300 dataset after being trained exclusively on TNG-Cluster (OOD test). 
        The dashed line denotes the 1:1 relation. 
        The kNN-Attn model (blue) achieves better alignment with the reference line and a lower RMSE (0.503) compared to the fine-tuned CNN (orange, RMSE = 0.729), which largely regresses to the mean and underestimates $t_{\mathrm{cool}}$ for both low and high values.}
    \label{fig:ood_overlay}
\end{figure}

We also investigate how model performance scales with training set size by varying the number of labeled examples on TNG300.
\begin{figure}[t]
    \centering
    \includegraphics[width=\linewidth]{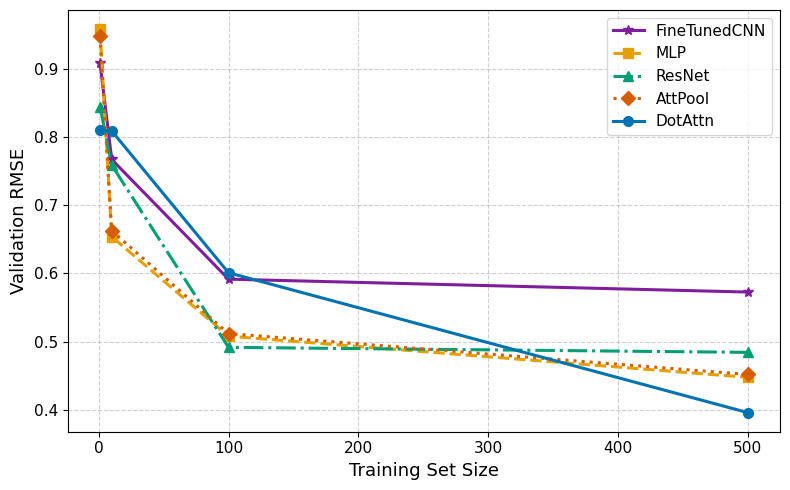}
    \caption[RMSE vs.\ training set size on TNG300]{
        Validation RMSE as a function of training set size on TNG300. 
        Each model was trained on subsets of size 1, 10, 100, and the full dataset. 
        kNN-Attn shows strong performance in low-data regimes.
    }
    \label{fig:data_scaling}
\end{figure}

\subsection{Structure in AstroCLIP Embeddings}

We begin our analysis by examining the structure of the learned AstroCLIP embedding space. As shown in the t-distributed Stochastic Neighbor Embedding (t-SNE) visualization in Figure~\ref{fig:tsne}, clusters with similar $t_{\mathrm{cool}}$ are naturally grouped together, indicating that the embedding captures physically meaningful information about the ICM. While both TNG-Cluster and TNG300 clusters inhabit the same global embedding space, they tend to populate distinct subregions, suggesting that domain-specific differences between the simulations—such as cluster populations—are preserved within the AstroCLIP representations despite being derived from a common encoder.

A natural question is why embeddings derived from a model trained on optical galaxy-scale images retain predictive power for X-ray cluster thermodynamics. We argue that this transferability arises from the low-level morphological features that Vision Transformers learn during pretraining. Regardless of wavelength, images of astrophysical systems encode structural information, such as surface brightness concentration, radial gradients, and azimuthal symmetry, that ViTs capture through their patch-based self-attention mechanism. In the context of galaxy clusters, these features are physically meaningful: cool-core clusters exhibit centrally peaked,  symmetric X-ray morphologies, while non-cool-core systems tend to be more disturbed and asymmetric. These morphological differences are precisely the kind of global, relational structure that ViTs are well-suited to encode. Although AstroCLIP was not trained to distinguish thermodynamic states, the features it extracts appear to be sufficiently general to capture the morphological signatures of ICM cooling. This is consistent with recent evidence that foundation model representations trained on large, diverse datasets develop transferable feature hierarchies applicable across domains \citep{bommasani_opportunities_2021}.

To quantify the utility of the learned embedding space, we evaluate its effectiveness on a downstream prediction task. We first implement a simple kNN baseline: for each cluster, we predict its $t_{\mathrm{cool}}$ as the mean of its four nearest neighbors in embedding space, discarding the one whose $t_{\mathrm{cool}}$ deviates most from the group average. This lightweight, training-free approach achieves RMSE values of 0.5517 on TNG300, 0.5882 on TNG-Cluster, and 0.8251 on the merged dataset. These findings confirm that the self-supervised AstroCLIP embeddings encode cooling-relevant physical structure and contain sufficient signal to support regression based on embedding similarity alone.

We also examine whether the embedding encodes halo mass or thermodynamic state by performing a t-SNE analysis on both the TNG300 and TNG-Cluster samples. In each dataset, clusters are divided into three quantile-based mass bins and into standard $t_{\mathrm{cool}}$ categories (SCC, WCC, NCC). As shown in Figures~\ref{fig:tng_300_massdist} and \ref{fig:tng_cluster_massdist}, the t-SNE components show substantial overlap between mass bins, indicating no strong separation by halo mass. In contrast, $t_{\mathrm{cool}}$ categories exhibit clear structure in the embedding space, with SCC and WCC systems forming more localized groups and NCC clusters displaying a broader, more dispersed distribution. These results demonstrate that the AstroCLIP embedding consistently captures the thermodynamic state of the ICM rather than halo mass, even in the high-mass regime.

\subsection{Supervised Regression with Pretrained Embeddings}

The embedding-based regressors achieve comparable performance across both datasets, with the Attn model slightly outperforming the MLP and ResNet in most scenarios (Table~\ref{tab:best_hparams}).

Incorporating local neighborhood information further improves predictive performance. As shown in Table~\ref{tab:best_hparams}, kNN-Attn achieves the lowest RMSE across both TNG300 and TNG-Cluster. This improvement is also evident in Figure~\ref{fig:dotattn}, where the predicted cooling times closely follow the one-to-one relation with the ground truth and exhibit relatively little scatter, including on the merged dataset. The strong agreement across all datasets indicates that leveraging local neighborhood information within the AstroCLIP embedding space yields more accurate and robust cooling-time predictions than directly regressing from the embeddings alone.

\subsection{OOD Evaluation}

To assess the generalization capability of our models, we perform an out-of-distribution (OOD) evaluation. In addition to embedding-based regressors, we train a CNN based on ResNet directly on raw $252 \times 252$ mock X-ray images. This CNN is fine-tuned from ImageNet-pretrained weights. Both the fine-tuned CNN and the kNN-Attn model are trained on the full TNG-Cluster dataset and evaluated on the TNG300 dataset, which represents a distinct simulation domain. We report RMSE as the evaluation metric to quantify model performance across domains. This setting is doubly challenging, as AstroCLIP was also pretrained on galaxy-scale images, not clusters.

Despite this, the kNN-Attn model maintains strong performance and outperforms all other methods, demonstrating robust generalization to a new simulation domain. As shown in Figure~\ref{fig:ood_overlay}, its predictions remain tightly correlated with the true $\log(t_{\mathrm{cool}})$, while the CNN exhibits larger residuals and scatter. This result is intuitive: CNNs operate by detecting local $3 \times 3$ correlations in image space, which is effective when features exhibit strong local spatial structure. However, AstroCLIP embeddings are global, high-level representations where such local operations may miss broader semantic relationships. In contrast, kNN-Attn uses a sparse attention matrix over a small set of semantically relevant neighbors, enabling global reasoning with far fewer trainable parameters.

\subsection{Performance as a Function of Training Data}



We also investigate how model performance scales with training set size by varying the number of labeled examples on TNG300. We compare five models: the four embedding-based regressors (MLP, ResNet, Attn, and kNN-Attn) and a fine-tuned ResNet18 CNN trained on raw 252×252 X-ray images, taken directly from \citet{sadikov_galaxy_2025} and trained on TNG300. Each model is trained on subsets of size 1, 10, 100, and the full dataset of clusters, and evaluated on a fixed validation set.

Figure~\ref{fig:data_scaling} shows that validation RMSE decreases for all models as the number of labeled training examples increases. Across all training set sizes, the embedding-based approaches consistently outperform the fine-tuned CNN, particularly in the low-data regime, demonstrating the strong transferability of AstroCLIP representations. While the MLP, ResNet, and Attn models exhibit similar performance, kNN-Attn continues to improve with increasing training data and achieves the lowest validation RMSE when trained on the full dataset. These results highlight both the data efficiency of pretrained AstroCLIP embeddings and the additional benefit of incorporating neighborhood information through attention.

\section{Discussion}\label{sec:discussion}
In this work, we demonstrate that, despite being trained on galaxies rather than galaxy clusters, AstroCLIP generalizes well to the cluster regime. Although galaxy clusters represent a distinct astrophysical system, their X-ray emission, like the optical emission of galaxies, exhibits characteristic morphological features. These features are likely responsible, at least in part, for AstroCLIP's ability to encode information related to cluster thermodynamic state. For example, Figures 3 and 4 show that clusters with different \(t_{\rm cool}\) values preferentially occupy distinct regions of the embedding space, whereas no comparable separation is observed between different mass bins. This suggests that the predictive power of the embeddings is not driven primarily by cluster mass, but instead by morphological features associated with cooling state, such as central surface-brightness concentration, radial gradients, and asymmetries in the X-ray emission.

Naturally, this raises the question of why the AstroCLIP foundation model remains so performant despite having been trained on optical images of galaxies rather than X-ray images of galaxy clusters. One possibility is that the Vision Transformer in AstroCLIP learns generic structural representations that transfer across these domains, including emission concentration, radial gradients, symmetry, and extended versus compact structure. These features are not unique to galaxies and can also encode physically meaningful differences in the X-ray morphology of galaxy clusters. Determining which of these morphological features contribute most strongly to the learned representation would require a dedicated interpretability analysis and represents an interesting direction for future work.




As previously discussed in this paper, several other works such as \cite{chadayammuri_ergo-ml_2026} and \cite{sadikov_galaxy_2025} have investigated using machine learning to predict galaxy cluster properties such as $t_{cool,0}$. While both papers use mock observations originating from IllustrisTNG data of galaxy clusters, they differ in their approaches to how the training data is used to infer physical properties. \cite{sadikov_galaxy_2025} trained a ResNet on the mock images themselves while \cite{chadayammuri_ergo-ml_2026} first train a contrastive learning algorithm to build a representative embedding space from the mock images before training a simple regressor to estimate cluster parameters. For comparison, the CNN model of \cite{sadikov_galaxy_2025} achieves a mean percentage error of \(1.8\%\) for the central cooling time, while the plane-based regression of \cite{chadayammuri_ergo-ml_2026}, applied to a UMAP representation obtained through contrastive learning, yields a mean fractional error of \(-2.87\pm3.27\%\) for t$_\mathrm{cool}$. These results similarly highlight the predictive information encoded in the learned representations. Both papers demonstrate that machine learning is capable of extracting physical parameters from mock galaxy cluster images.

While we have demonstrated a similar ability, our approach differs in multiple significant ways: we use a foundation model, AstroCLIP, to embed the clusters, we apply a downstream model to infer $t_{cool,0}$, and we train on a larger set of representative clusters. In particular, we demonstrate that AstroCLIP, a foundation model trained on photometric and spectroscopic optical observations of individual galaxies, is capable of embedding X-ray images into a physically meaningful embedding space. By using a pre-trained foundation model, we only train a kNN-Attn model using the X-ray cluster images. This juxtaposes the aforementioned studies which relied purely on the X-ray cluster images to learn the entire representation. Moreover, this carries several meaningful benefits. We demonstrate that our methodology extends cleanly to out-of-distribution data and is able to generalize to datasets with a small amount of labeled data.

These distinctions are particularly important when considering applying these methods to real observations where it is effectively impossible to construct large homogeneous labeled datasets. We demonstrate that using a foundation model such as AstroCLIP, downstream machine learning algorithms can be tuned on relatively small sets of labeled data. 

Interestingly, the kNN-Attn network achieves substantially better performance when trained independently on TNG300 and TNG-Cluster (\(\mathrm{RMSE}=0.23{-}0.24\)) than when trained simultaneously on the combined TNG300 and TNG-Cluster sample (\(\mathrm{RMSE}=0.59\)). The embedding distributions provide some insight into this difference. As shown in Figure 2, although the TNG300 and TNG-Cluster populations overlap in the AstroCLIP embedding space, they preferentially occupy different regions. The poorer performance on the combined sample may therefore indicate that the relationship between the AstroCLIP representation and \(t_{\rm cool,0}\) is not identical across the two cluster populations. Determining the origin of this difference would require further investigation.

Nevertheless, the out-of-distribution experiment demonstrates that the learned representation retains predictive power across the two datasets. When trained on TNG-Cluster and evaluated on TNG300, kNN-Attn achieves an RMSE of \(0.50\), compared with \(0.73\) for the CNN trained directly on the X-ray images. This suggests that the pretrained AstroCLIP representation provides some robustness to differences between the cluster populations. However, TNG300 and TNG-Cluster share the same underlying IllustrisTNG galaxy-formation model and subgrid physics, and therefore this experiment does not test generalization across fundamentally different physical models. A more stringent test will require applying the framework to clusters drawn from independent cosmological simulations, such as EAGLE \citep{schaye_eagle_2015, crain_eagle_2015} and The Three Hundred \citep{cui_three_2018}, which adopt different implementations of baryonic physics and feedback.

While we have demonstrated that the framework presented in this work can accurately predict \(t_{\rm cool,0}\) from realistic mock X-ray observations, it remains to be determined whether this performance extends to real observations. The transition from simulated to observed clusters represents a substantially more challenging domain shift than those considered here. Although our mock observations incorporate many of the instrumental and observational effects expected in Chandra data, differences between the simulated and observed cluster populations, as well as variations in exposure time, background, angular resolution, and redshift, may affect the learned representations and subsequent predictions.

An important next step is therefore to apply the framework to a sample of observed clusters for which \(t_{\rm cool,0}\) has been independently measured. Such a test would allow the predicted cooling times to be directly compared with values derived from traditional spectroscopic analyses and provide a critical assessment of the simulation-to-observation domain shift. Future work will first extend our out-of-distribution tests to clusters drawn from independent cosmological simulations before ultimately applying the framework to real Chandra observations.

If successful, this would provide a path toward estimating cluster cooling properties from X-ray imaging alone, enabling the characterization of much larger cluster samples from current and future X-ray surveys such as eROSITA.
\section{Conclusion} \label{sec:conclusion}

Estimating the cool-core status of galaxy clusters is essential for understanding the thermodynamics of the ICM and the regulation of cooling by AGN feedback. A variety of approaches have been developed, including observational diagnostics, simulation-based proxies, and machine learning techniques, each differing in their physical interpretability, data requirements, and generalizability. Observational studies often classify clusters based on thresholds in $t_{\mathrm{cool}}$ \citep[e.g.,][]{hudson_impact_2010}, or use entropy-based proxies such as the core entropy $K_0$ \citep{cavagnolo_intracluster_2009}. Other works employ morphological indicators such as the surface brightness concentration (CSB) \citep[e.g.,][]{santos_revisiting_2008, andrade-santos_origin_2018}. While effective, these methods typically require high-resolution X-ray spectroscopy and rely on hand-crafted thresholds.

Simulation-based studies have provided further insight into cool-core evolution and the physical processes driving core disruption, including the role of mergers \citep{hahn_revisiting_2017}, and have compared simulated cool-core statistics with observations in cosmological runs such as Illustris and IllustrisTNG \citep[e.g.,][]{barnes_role_2018, truong_cool-core_2023}. Recently, \citet{sadikov_galaxy_2025} applied unsupervised clustering, deep learning regression, and simulation-based inference to mock \textit{Chandra} X-ray images from the $z=0$ snapshot of the IllustrisTNG simulations, successfully predicting multiple cool-core diagnostics—including central $t_{\mathrm{cool}}$, entropy excess, and concentration parameter—with high accuracy, thereby demonstrating the potential of machine learning for large X-ray surveys such as \textit{eROSITA}. However, such simulation-based analyses often depend on access to full three-dimensional gas profiles, which are not available observationally.

In this work, we propose a complementary approach that combines the strengths of traditional and machine learning methods. We employ the contrastive learning model AstroCLIP to obtain physically meaningful, domain-robust embeddings of cluster X-ray images, and then apply supervised regression to predict $\log(t_{\mathrm{cool}})$ directly from these embeddings. This strategy is data-efficient, generalizes across simulated domains, and avoids the need for full gas profiles or manually defined thresholds.

Using projections from both TNG300 and the high-resolution TNG-Cluster suite, we evaluated four supervised regression models and introduced a hybrid kNN-Attn approach that combines neighbourhood retrieval with attention-based weighting. This model achieved the lowest RMSE across both datasets (0.24 for TNG300 and 0.23 for TNG-Cluster) and demonstrated strong generalization in an out-of-distribution setting, outperforming a fine-tuned convolutional neural network (RMSE 0.50 versus 0.73).

We showed that clusters with similar cooling times occupy nearby regions in embedding space, and that simple nearest-neighbour approaches already yield reasonable predictions of $t_{\mathrm{cool}}$. t-SNE visualizations further reveal physically meaningful structure correlated with $t_{\mathrm{cool}}$, indicating that the learned representation captures thermodynamic information despite being trained on a different astrophysical domain.

These results highlight two key findings. First, pretrained foundation model embeddings can transfer across domains and retain physically meaningful structure relevant to galaxy cluster thermodynamics. Second, leveraging local structure in embedding space through nearest-neighbour retrieval and attention provides a robust and data-efficient approach to regression, particularly under domain shift.
Although this study relies on AstroCLIP embeddings, the proposed framework is not tied to a specific foundation model and can be readily extended to future representation-learning models. Recent developments such as AION-1 \citep{parker_aion-1_2025}, an omnimodal foundation model trained on diverse astronomical data modalities, may provide even richer astrophysical representations than those used here. Investigating whether AION-derived embeddings improve cooling-time regression, cool-core classification, and cross-domain generalization represents a promising direction for future work.

This framework offers a promising pathway toward scalable, spectroscopically independent inference of cluster cooling properties. Our approach is directly applicable to large, heterogeneous datasets and all-sky X-ray surveys such as \textit{eROSITA}.

\section*{Data Availability}

The inclusion of a Data Availability Statement is a requirement for articles published in MNRAS. Data Availability Statements provide a standardised format for readers to understand the availability of data underlying the research results described in the article. The statement may refer to original data generated in the course of the study or to third-party data analysed in the article. The statement should describe and provide means of access, where possible, by linking to the data or providing the required accession numbers for the relevant databases or DOIs.

\section*{acknowledgements}
JHL acknowledges funding support from the Canada Research Chairs Program, as well as the Natural Sciences and Engineering Research Council of Canada (NSERC) through the Discovery Grant, Accelerator Supplement programs and the Arthur B. McDonald Fellowship. 



\bibliographystyle{mnras}
\bibliography{example} 





\end{document}